\documentclass[floatfix,aps,prl,twocolumn,showpacs,superscriptaddress,preprintnumbers]{revtex4-1}

\usepackage[english]{babel}
\usepackage{bm}    
\usepackage{epsfig}
\usepackage{amsbsy,amssymb} 
\usepackage{graphicx}
\usepackage{array}
\usepackage{amsmath}
\usepackage{multirow}
\usepackage{natbib}

\newcommand\new{\newcommand}         

\newcolumntype{M}{>{\centering\arraybackslash}m{\dimexpr.33\linewidth-2\tabcolsep}}

\def\beq{\begin{equation}}   
\def\eeq{\end{equation}}
\def\bea{\begin{eqnarray}}  
\def\eea{\end{eqnarray}} 
\newcommand{\bite}{\begin{itemize}}
\newcommand{\eite}{\end{itemize}}

\def\GAT{\textrm{GA}_{T}}

\new{\eV}         {{\ifmmode {\mathrm{ eV}}\else ${\mathrm{ eV}}$\fi}}
\new{\MeV}        {{\ifmmode {\mathrm{ MeV}}\else ${\mathrm{ MeV}}$\fi}}
\new{\MeVc}       {{\ifmmode {\mathrm{ MeV}}/c\else ${\mathrm{ MeV}}/c$\fi}}
\new{\MeVcc}      {{\ifmmode {\mathrm{ MeV}}/c^2\else ${\mathrm{ MeV}}/c^2$\fi}}
\new{\GeV}        {{\ifmmode {\mathrm{ GeV}}\else ${\mathrm{ GeV}}$\fi}}
\new{\GeVc}       {{\ifmmode {\mathrm{ GeV}}/c\else ${\mathrm{GeV}}/c$\fi}}
\new{\GeVcc}      {{\ifmmode {\mathrm{ GeV}}/c^2\else ${\mathrm{GeV}}/c^2$\fi}}
\new{\TeV}        {{\ifmmode {\mathrm{ TeV}}\else ${\mathrm{ TeV}}$\fi}}

\new{\Mh}         {{\ifmmode M_{\mathrm{ H}}
                    \else $M_{\mathrm{H}}$\fi}}

\new{\Mz}         {{\ifmmode M_{\mathrm{Z}}
                    \else $M_{\mathrm{Z}}$\fi}}
\new{\Mzsq}       {{\ifmmode M^2_{\mathrm{ Z}}
                    \else $M^2_{\mathrm{Z}}$\fi}}

\new{\as}[1]      {{\ifmmode\alpha^{#1}_s
                    \else$\alpha^{#1}_s$\fi}}
\new{\asx}[1]      {{\ifmmode a^{#1}_s
                    \else $a^{#1}_s$\fi}}
\new{\asb}[1]     {{\ifmmode\overline{\alpha}^{#1}_s
                    \else $\overline{\alpha}^{#1}_s$\fi}}
\new{\asmz}       {{\ifmmode\alpha_s(\Mzsq)
                    \else $\alpha_s(\Mzsq)$\fi}}
\new{\lqcd}       {{\ifmmode\Lambda_{\mathrm{ QCD}}
                    \else $\Lambda_{\mathrm{ QCD}}$\fi}}

\def\Gosam{{{\sc GoSam}}}

\def\samurai{{{\sc samurai}}}
\def\Sherpa{{{\sc Sherpa}}}

\def\C++{{{\sc c++}}}

\def\OneLoop{{{\sc OneLoop}}}

\def\Ninja{{{\sc Ninja}}}
\def\Cpp{{{\sc C++}}}

\def\Amegic{{{\sc Amegic}}}

\newcommand{\mpi}{Max-Planck-Institut f\"ur Physik, F\"ohringer Ring 6, 80805 M\"unchen, Germany}
\newcommand{\padova}{Dipartimento di Fisica e Astronomia, Universit\`a di Padova, and INFN  \\                                    
Sezione di Padova, via Marzolo 8, 35131 Padova, Italy}
\newcommand{\cuny}{Physics Department, New York City College of Technology, The City University of New York, 300 Jay Street Brooklyn, NY 11201, USA}
\newcommand{\cunygc}{The Graduate School and University Center, The City University of New York,
365 Fifth Avenue, New York, NY 10016, USA}

\begin{document}


\title{NLO QCD corrections to Higgs boson production in association \\ 
         with a top quark pair and a jet}

\author{H.~van~Deurzen}
\email{hdeurzen@mpp.mpg.de}
\affiliation{\mpi}
\author{G.~Luisoni}%
\email{luisonig@mpp.mpg.de}
\affiliation{\mpi}
\author{P.~Mastrolia}
\email{pierpaolo.mastrolia@cern.ch}  
\affiliation{\mpi} 
\affiliation{\padova}
\author{E.~Mirabella}
\email{mirabell@mpp.mpg.de}  
\affiliation{\mpi}
\author{G.~Ossola} 
\email{gossola@citytech.cuny.edu}
\affiliation{\cuny \\ \cunygc}
\author{T.~Peraro}
\email{peraro@mpp.mpg.de}  
\affiliation{\mpi}

\begin{abstract}
We present the calculation of the cross section for Higgs boson
production in association with a top quark pair plus one jet, at
next-to-leading-order (NLO) accuracy in QCD. All mass dependence is retained without recurring to any approximation. After including the complete NLO QCD corrections, we
observe a strong reduction in the scale dependence of the result. We also show distributions for the 
invariant mass of the top quark pair, with and without the additional jet, and for  the transverse momentum and the pseudorapidity of the Higgs boson. Results for the virtual contributions are obtained with a novel reduction approach based on integrand decomposition via Laurent expansion, as implemented in the library {\Ninja}. Cross sections and differential distributions are obtained with an automated setup which combines the  {\Gosam} and  {\Sherpa} frameworks.
\end{abstract}

\pacs{}

\maketitle

\section{Introduction}
\label{Sec:intro}

The evidence of the existence of a new particle of mass between 125
and 126 GeV, initially reported about one year ago by the ATLAS
and CMS collaborations~\cite{Aad:2012tfa,Chatrchyan:2012ufa}, has been
confirmed with very high confidence level by more recent analyses,
thus providing more stringent arguments in favor of the validity of the electroweak symmetry breaking mechanism.
It is interesting to observe that all the analyses performed so far are in good agreement with the hypothesis
that the new particle is the Higgs boson predicted by the
Standard Model (SM). Indeed, rates and distributions 
are compatible with the assumption that the new particle is a scalar
that couples to other SM particles with a strength proportional to
their mass~\cite{CMS:yva,Aad:2013wqa, Aaltonen:2013kxa}. 
Accurate predictions 
are necessary and will play a crucial role for the complete
determination of the nature of the Higgs boson~\cite{Heinemeyer:2013tqa},
in particular to shed light on the structure of its couplings to the other particles.

The production rate for a Higgs boson associated with a top-antitop pair ($t \bar t H$) is particularly interesting in this context, since it is directly proportional to the SM Yukawa coupling of the Higgs boson to the top quark. 
The study of differential observables and distributions will bring
information on the coupling structure and on the parity of the Higgs particle~\cite{Frederix:2011zi, Degrande:2012gr}. 

The difficulties related to the analysis of the $t \bar t H$ channel
are well known. The combined production of three heavy particles requires a large center-of-mass energy for the initial partons, which is strongly suppressed by parton 
distribution functions. Furthermore, additional difficulties are represented by the presence of various challenging
backgrounds and by the complexity of the final state, which make its
kinematic reconstruction far from straightforward~\cite{Artoisenet:2013vfa}.

\begin{figure}[t]
\begin{center}
\includegraphics[width=6cm]{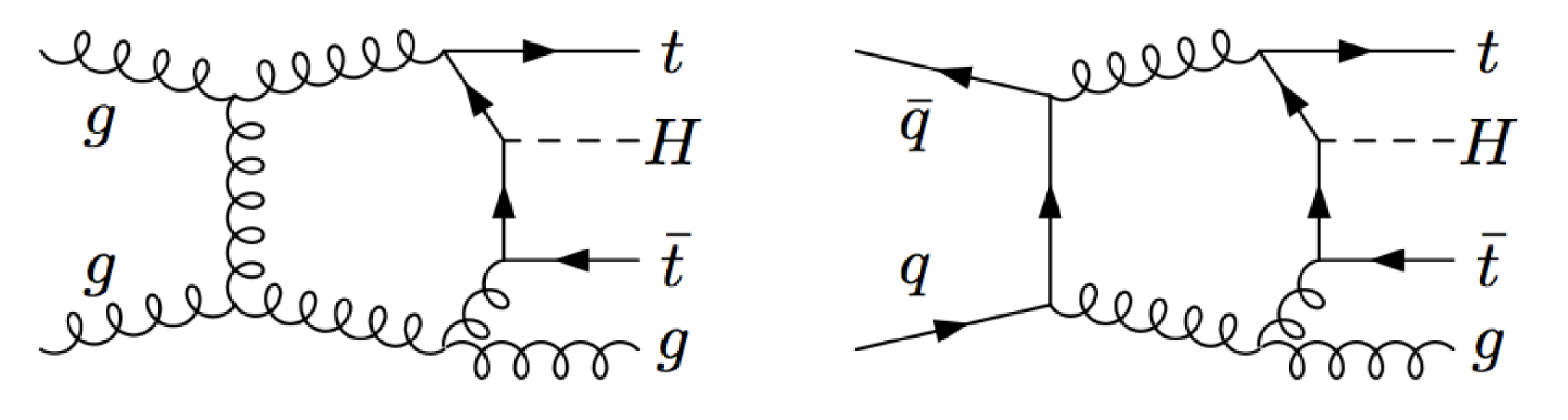}
\caption{Sample of one-loop diagrams contributing to the NLO corrections
to $g g \to t {\bar t} H g $ and $q {\bar q} \to t {\bar t} H g$.}
\label{hexagons}
\end{center}
\end{figure}

At the parton level, the $t \bar t H$ production at next-to-leading order (NLO) in QCD has been known 
for some time~\cite{Beenakker:2001rj,Beenakker:2002nc,Dawson:2002tg,
Dawson:2003zu,Dittmaier:2003ej}. More recently, this process has been employed in a number of  studies, motivated by the new analyses performed at the LHC~\cite{Plehn:2009rk, Frederix:2011zi, Degrande:2012gr, Artoisenet:2013vfa}. 

In this letter, we present the complete NLO QCD corrections to
the process $ pp \to t \bar t H + 1$ jet  ($t \bar t H j$) at the LHC. Examples of contributing one-loop diagrams are depicted in Fig.~\ref{hexagons}. 
We illustrate the outcome of our calculation by showing the total cross section, and a selection of differential distributions.

The goal of the considered calculation is twofold. 
On the one hand, it is important for the phenomenological analyses at
the LHC, in particular for  the high-$p_T$ region, where the presence of the additional jet can be
sensibly relevant. 
On the other hand, $t \bar t H j$  constitutes 
 the first application of a novel reduction
algorithm for the evaluation of one-loop amplitudes, which strengthens the performances of the integrand
decomposition \cite{Mastrolia:2012bu}, in particular in the presence of massive particles.
\section{Computational set-up}
\label{Sec:calc}

In perturbation theory, computations at the NLO accuracy require, aside from the evaluation of
leading-order (LO) contributions,  the calculation of both virtual and
real-emission corrections.
The Born and the real emission matrix elements are computed using {\Sherpa}~\cite{Gleisberg:2008ta} and the library \Amegic~\cite{Krauss:2001iv}, which implements the Catani-Seymour dipole 
formalism~\cite{Catani:2002hc, Gleisberg:2007md}.  {\Sherpa}  also performs the integration over the phase space and the analysis.  The virtual corrections are 
generated with the {\Gosam} package~\cite{Cullen:2011ac}, which combines automated diagram generation 
and algebraic manipulation~\cite{Nogueira:1991ex, Vermaseren:2000nd, Reiter:2009ts, Cullen:2010jv, Kuipers:2012rf}  with $d$-dimensional integrand-level reduction 
 techniques~\cite{Ossola:2006us, Ossola:2007bb, Ellis:2007br, Ossola:2008xq,  Mastrolia:2008jb, Mastrolia:2010nb, Heinrich:2010ax}. 
 The master integrals (MIs) are computed using {\OneLoop}~\cite{vanHameren:2010cp}.  
The code generated by {\Gosam} is linked to {\Sherpa} by means of the Binoth
Les Houches Accord (BLHA)~\cite{Binoth:2010xt} interface, which uses 
a system of {\it order} and {\it contract} files and allows for a direct communication between the two codes at running time.   
The same setup  has been recently employed for the computation of NLO QCD corrections to $p p \to H j j$~\cite{vanDeurzen:2013rv}  and  $p p \to H j j j $~\cite{Cullen:2013saa} (for the latter, in combination with {\sc MadDipole/MadEvent}~\cite{Maltoni:2002qb,Frederix:2008hu,Frederix:2010cj}) and also for the analysis of the $t \bar t$ forward-backward asymmetry~\cite{Hoeche:2013mua}.

For  $t \bar t H j$ production, the basic partonic processes identified
by the {\Sherpa}-{\Gosam} contract file are:
 \beq
    q \, \bar q \,  \to \, t \, \bar t \, H  \, g  \, ,   \qquad        
    g \,  g  \,  \to \, t \, \bar t \, H  \, g \, ,   
 \label{Eq:Pproc}
 \eeq
while the remaining subprocesses can be obtained by proper crossings.
 The ultraviolet, the infrared, and the collinear
 singularities are regularized using dimensional reduction. 
 The renormalization conditions are fixed along the lines
 of~\cite{Beenakker:2002nc,Dawson:2003zu}, where 
 the top mass is renormalized on-shell, while  the strong coupling is 
 renormalized in the $\overline{\mbox{MS}}$ scheme, decoupling 
 the top quark from the running.  
 In the case of LO [NLO] contributions, we describe the running of the 
 strong coupling constant with one-loop [two-loop] accuracy. 
The wave functions
of the gluon and of the quarks are renormalized on-shell, {\it i.e.} 
the corresponding renormalization constants cancel 
the external self-energy corrections exactly.

The virtual amplitudes of $t {\bar t}Hj$ have been decomposed in terms
of MIs using for the first time 
the {\it integrand reduction via Laurent expansion}~\cite{Mastrolia:2012bu}, implemented in the {\Cpp} library \Ninja.
This new algorithm exploits the complete knowledge of the analytic
expression of the integrand and of the residues at the multiple cut  
to ameliorate the determination of the coefficients of the MIs
with respect to the canonical integrand reduction~\cite{Ossola:2006us}.
Elaborating on  the techniques introduced in~\cite{Forde:2007mi,Kilgore:2007qr,Badger:2008cm},
the series expansion combined with the integrand decomposition lowers  
the computational load and improves the  accuracy of
the results. 
Within this new algorithm, the sampling of the numerators and the
subtractions of the higher-point residues, characterizing the 
{\it triangular} system-solving approach of the original integrand-reduction procedure, are avoided.
Instead, the series expansion allows for a {\it diagonal} system-solving 
strategy, where the polynomial subtractions of the residues,
when needed, are replaced by universal correction terms 
which have to be added to the coefficients of the Laurent series.
These universal corrections, required only for the determination of
the coefficients of 2-point and 1-point MIs, are obtained, once and
for all, from the expansions of the generic polynomial forms of the 
residues at the triple and double cuts.

\begin{figure}[t]
\begin{center}
\includegraphics[width=8cm]{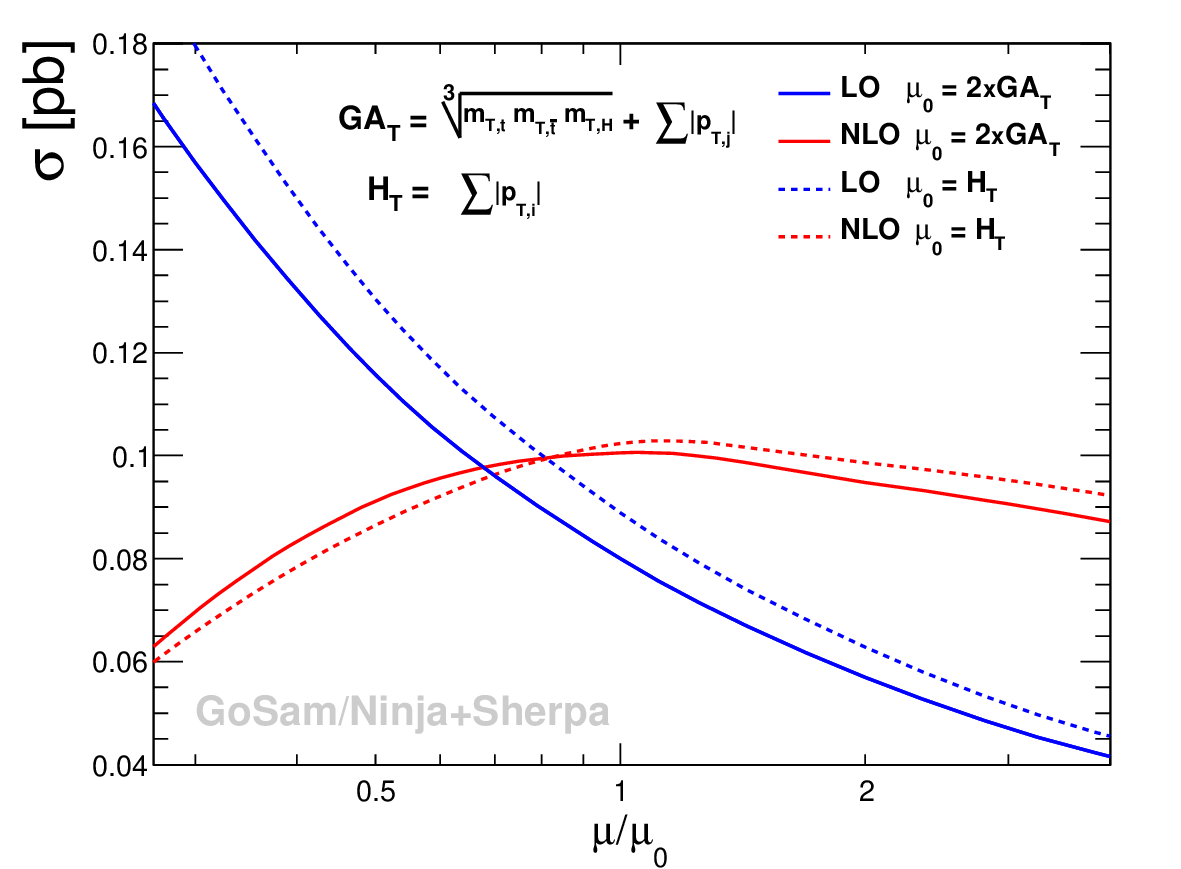} 
\caption{Scale dependence of the total cross section at LO and NLO.}
\label{Fig:scalevars}
\end{center}
\end{figure}

The {\Ninja} library, which has been interfaced to {\Gosam}, 
implements the integrand reduction via Laurent expansion using 
a semi-analytic algorithm.  The coefficients of the Laurent 
expansion of a generic integrand  are efficiently computed by performing a 
{\it polynomial division} between the numerator and the set of uncut denominators~\cite{Mastrolia:2012bu}.
      
The calculation of the NLO virtual corrections performed with  {\Ninja} has been checked using the 
independent reduction algorithm
implemented in the 
library {\samurai}~\cite{Mastrolia:2010nb}. 
We verified the agreement of
the virtual corrections obtained with the two reduction procedures in ten thousand
phase-space points. 
The values of double and the single poles, for each individual subprocess, conform to the universal singular behavior of
dimensionally regulated one-loop amplitudes~\cite{Catani:2000ef}.
Our results fulfill gauge invariance, verified 
through the vanishing of the amplitudes when substituting the polarization
vector of one or more gluons with the corresponding 
momentum. 

The {\Ninja} reduction algorithm proved
to be numerically more efficient and stable.  
In fact, for the highly non-trivial process under consideration, 
only a small set of phase-space points, of the order of few per mill,
were detected as unstable. 
All these points have been recovered using the tensorial reduction provided 
by {\sc Golem95}~\cite{Binoth:2008uq,Cullen:2011kv}, thus avoiding the necessity of higher precision routines, which are extremely time consuming.

\begin{figure}[h]
\begin{center}
\includegraphics[width=7.5cm]{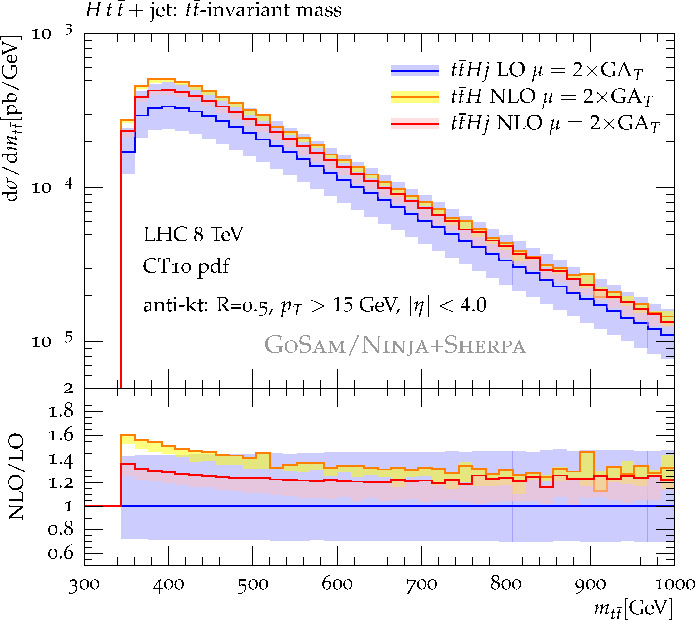}
\caption{Invariant mass distributions of the $t\bar{t}$-pairs for $t \bar t H$ and $t \bar t H j$ at NLO relative to the $t \bar t H j$ at LO for $\mu=2\times\GAT$.}
\label{Fig:inv1}
\end{center}
\end{figure}

The time required for the computation of the full color- and helicity-summed amplitudes in one phase-space point is about  $2.5$ seconds. The numerical values of the one-loop amplitudes for the two
partonic processes listed in Eq.~(\ref{Eq:Pproc})  in a non-exceptional phase-space point 
are collected in the Appendix.  

In view of the later comparison between the processes $pp \to t {\bar t} H$
and $pp \to t {\bar t} Hj$ at NLO QCD accuracy, 
we also used the {\Gosam/\Ninja}+{\Sherpa} framework to compute the cross section for 
$t {\bar t} H$ production.  
We found excellent agreement with the results presented in
Refs.~\cite{Hirschi:2011pa,Frederix:2011zi}.
\section{Numerical results}
\label{Sec:resu}

In the following, we present results for the integrated cross section
for a center-of-mass energy of $8$ TeV.  The mass of the Higgs boson
is set to $m_H=126$ GeV and the top quark mass is set to $m_t=172.5$
GeV. The parameters of the electroweak sector are fixed by setting $M_W=80.419$ GeV,
$M_Z=91.1876$ GeV and $\alpha_{EW}^{-1}=132.50698$. 

To cluster the jets we use the {\tt antikt}-algorithm implemented in
{\sc FastJet}~\cite{Cacciari:2005hq,Cacciari:2008gp,Cacciari:2011ma}
with radius $R=0.5$, a minimum transverse momentum of
$p_{T,jet}>15$ GeV and pseudorapidity $|\eta|<4.0$. The LO cross
sections are computed with the LO parton-distribution functions cteq6L1~\cite{Pumplin:2002vw},
whereas at NLO we use CT10~\cite{Lai:2010vv}.

In order to study the scale dependence of the total cross section, we
employ two different choices of the renormalization and factorization
scales $\mu_{R} = \mu_{F} = \mu_{0}$, namely $\mu_0 = H_T$ and $\mu_0 = 2\times\GAT$
with
\begin{align}
H_T & = \sum_{ 
\tiny{
\begin{array}{cc} \mbox{\tiny final }  \label{eq:ht} \\
\mbox{ \tiny states } f  \end{array} }
} 
\left | p_{T,f}\right | \; , \\
\GAT & =\sqrt[3]{m_{T,H}\, m_{T,t}\, m_{T,\bar t}} + \sum_{ \tiny{ \mbox{\tiny jets } j }} |p_{T,j}| \label{eq:ga} \, .
\end{align}

\begin{figure}[t]
\begin{center}
\includegraphics[width=7.5cm]{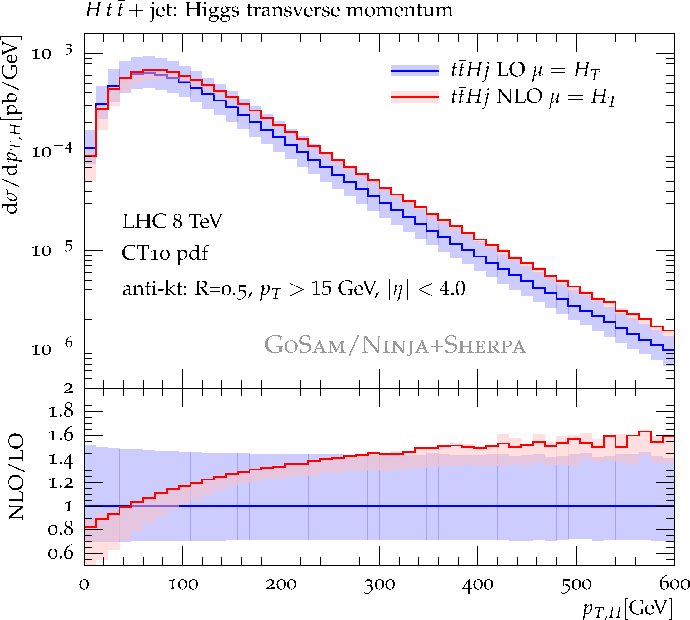} 
\caption{Transverse momentum distribution of the Higgs boson at LO and NLO for $\mu=H_T$.}
\label{Fig:pth}
\end{center}
\end{figure}

\begin{figure}[t]
\begin{center}
\includegraphics[width=7.5cm]{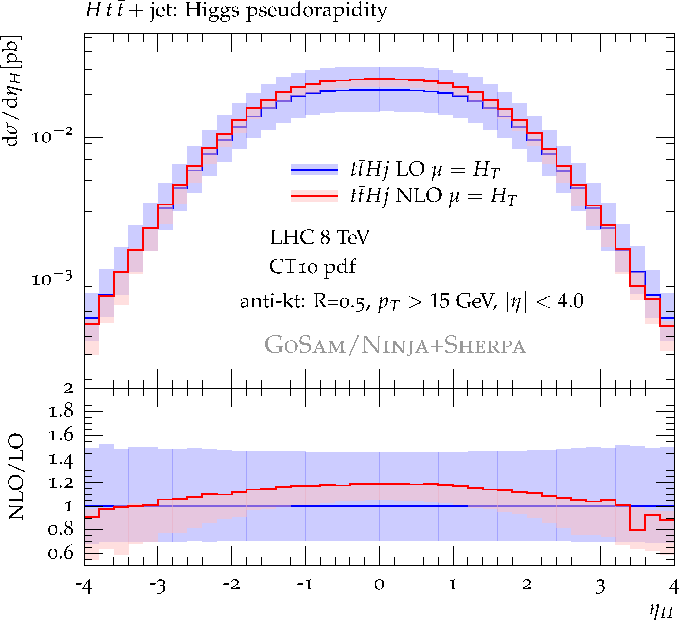} 
\caption{Pseudorapidity $\eta$ of the Higgs boson at LO and NLO accuracy for $\mu=H_T$.}
\label{Fig:etah}
\end{center}
\end{figure}

Within this setup, for the two scale choices, we obtain the total LO
and NLO cross sections 
reported in Table~\ref{Tab:xs}.

\begin{table}[h]
\begin{tabular}{MMM}
\hline 
\hline
 {\sc Central Scale} & $\sigma_{LO}$ [fb] & $\sigma_{NLO}$ [fb] \\
\hline \\ [-1.7ex]
2$\times\GAT$ \hspace{10pt} & $80.03^{+35.64}_{-23.02}$ \hspace{10pt} & $100.6^{+0.00}_{-9.43}$  \\ 
\\ 
$H_T$       \hspace{10pt} & $88.93^{+41.41}_{-26.13}$  \hspace{10pt}
& $102.3^{+0.00}_{-15.82}$  \\
 \\ [-1.7ex]
\hline
\hline
\end{tabular}
\caption{Total cross section for $t \bar t H j$  for different choices of the central scale at LO and NLO.}
\label{Tab:xs}
\end{table}
The scale dependence of the
total cross section, depicted in Fig.~\ref{Fig:scalevars}, is strongly
reduced by the inclusion of the NLO contributions. It is worthwhile to
notice that both choices for the central value of the scale provide an
adequate description, being close to the physical scale of the
process.

In Fig.~\ref{Fig:inv1}, we compare the distributions for the invariant mass of the top quark pair in $pp \to t {\bar t} H j$ at LO and NLO with the NLO curve for $pp \to t {\bar t} H$. For $t {\bar t} H j$,  going from LO to NLO accuracy, we observe an increase in the distribution by 20--35\% over the full kinematical range. On the other hand, when comparing the NLO $t {\bar t} H$ prediction with the NLO $t {\bar t} H j$ curve, the cross section decreases due to the presence of the additional jet which takes away energy from the $t\bar{t}$ system. This is particularly evident near the $t\bar{t}$ production threshold, while for high values of the $t {\bar t}$ invariant mass the two NLO curves get closer. The scale for this comparison is set to $\mu=2\times\GAT$.

In Fig.~\ref{Fig:pth} and Fig.~\ref{Fig:etah}, we display the distributions of the transverse momentum $p_T$ and the pseudorapidity $\eta$ of the Higgs boson, respectively. Each 
plot contains the distributions at LO and NLO accuracy, for a value of the scale set to  $\mu=H_T$. 
The NLO corrections are particularly important for high values of the $p_T$, which are the kinematical regions involved in the boosted analyses~\cite{Butterworth:2008iy, Plehn:2009rk}.

These distributions show the potential of the framework obtained combining {\Gosam/\Ninja} with {\Sherpa}, which can be successfully used to compute NLO predictions for multi-leg processes 
involving massive particles.
Moreover they shed some light on the impact of further jet activity in $p p \to t \bar t H$, one of the most important processes for the direct determination of the coupling of the Higgs boson to fermions.
The NLO QCD corrections  reduce the scale uncertainty and their numerical impact  can be sizable. Therefore they  could be helpful for an accurate simulation of the signal in the experimental searches looking for Higgs production in association with a top-antitop pair at the LHC.

\medskip

\paragraph*{Acknowledgments --} We thank all the other members of the {\Gosam} project for collaboration on
the common development of the code.  We also thank  Jan Winter for interesting discussions.
The work of  H.v.D., G.L., P.M., and T.P. was  supported by the Alexander von
Humboldt Foundation, in the framework of the Sofja Kovaleskaja Award Project
``Advanced Mathematical Methods for Particle Physics'', endowed by the German
Federal Ministry of Education and Research.  G.O. was supported in part by the 
National Science Foundation under Grant PHY-1068550.  Computing
resources were provided by the CTP cluster of the New York City
College of Technology.

\bigskip

\appendix*

\section{Benchmark phase-space point}
\label{App:Bench}

In this appendix we collect  numerical  results for the renormalized virtual contributions
to the processes~(\ref{Eq:Pproc}), in correspondence to the phase-space point in Table~\ref{pspoint}.
The results are collected in Table~\ref{benchmark} and are computed using dimensional reduction.  
The  coefficients $a_i$ are  defined as follows:
\bea
\frac{a_{-2}}{\epsilon^2} +  \frac{a_{-1}}{\epsilon} + a_0   \equiv \frac{  2 \mathfrak{Re} \left  \{ \mathcal{M}^{\mbox{\tiny tree-level} \ast } \mathcal{M}^{\mbox{\tiny one-loop}  }   \right  \}  }{
( \alpha_s / 2 \pi) \left |  \mathcal{M}^{\mbox{\tiny tree-level}} \right |^2 \nonumber
 }  \, .
\label{Eq:AI}
\eea
The reconstruction  of the renormalized pole can be checked against the value of 
$a_{-1}$ and $a_{-2}$ obtained by the universal singular behavior 
 of the dimensionally regularized  one-loop amplitudes~\cite{Catani:2000ef}, while the precision of the finite parts is 
 estimated by re-evaluating the amplitudes for a set of momenta rotated by an arbitrary angle about the axis of collision.

	\begin{table*}[ptb]
		\centering
		\begin{tabular}{c c c c c}
		\hline \hline
			particle & $E$& $p_x$ & $p_y$ & $p_z$ \\ 
		\hline
		
$p_1$ &    250.00000000000000    &    0.0000000000000000 & 
0.0000000000000000   &     250.00000000000000 \\
$p_2$ &   250.00000000000000  &      0.0000000000000000 &
0.0000000000000000 &      -250.00000000000000 \\
$p_3$ & 177.22342332868467   &    -31.917865771774753 &
-19.543909461587205    &   -15.848571666570733 \\
$p_4$ &  174.89951284907735     &   13.440699620020803 &
24.174898117950033   &     -8.2771667589629576 \\
$p_5$ &    126.37478917634435   &     6.8355633672742222 &
-3.2652801590882752     &   6.0992096455298030 \\
$p_6$ &  21.502274645893632     &   11.641602784479652 &
-1.3657084972745175  &      18.026528780003872\\
	
		\hline \hline
		\end{tabular}
		\caption{Benchmark phase-space point for  $ t \bar t H j$ production}
		\label{pspoint}
	\end{table*}

\newcommand{\trule}{\rule[-1.5mm]{0mm}{5mm}}

\begin{table}[h]
\begin{tabular}{ l  r  r  } \hline \hline \trule
         & $q \bar q \to  t \bar t H g$      &  $g g  \to  t \bar t H g$    \\  
\hline  \trule
$a_0$     \qquad  & \underline{-80.40233}474207548              &   \underline{-45.697933}52903498 \\
$a_{-1}$  \qquad & \underline{-32.690291020}67917            & \underline{-35.9217497445}3633 \\
$a_{-2}$  \qquad & \underline{-5.66666666666}7901        & \underline{-9.00000000000}6723   \\

\hline
\hline 
\end{tabular}
\caption{Numerical results for the two subprocesses listed in
  Eq.~(\ref{Eq:Pproc}) evaluated at the phase-space point of
  Table~\ref{pspoint} for a scale $\mu=2 p_1\cdot p_2 = 500$ GeV. The accuracy of the result is indicated by
  the underlined digits.
} 
\label{benchmark}
\end{table}

\bibliographystyle{apsrev}
\bibliography{references.bib}

\begin{thebibliography}{54}
\expandafter\ifx\csname natexlab\endcsname\relax\def\natexlab#1{#1}\fi
\expandafter\ifx\csname bibnamefont\endcsname\relax
  \def\bibnamefont#1{#1}\fi
\expandafter\ifx\csname bibfnamefont\endcsname\relax
  \def\bibfnamefont#1{#1}\fi
\expandafter\ifx\csname citenamefont\endcsname\relax
  \def\citenamefont#1{#1}\fi
\expandafter\ifx\csname url\endcsname\relax
  \def\url#1{\texttt{#1}}\fi
\expandafter\ifx\csname urlprefix\endcsname\relax\def\urlprefix{URL }\fi
\providecommand{\bibinfo}[2]{#2}
\providecommand{\eprint}[2][]{\url{#2}}

\bibitem[{\citenamefont{Aad et~al.}(2012)}]{Aad:2012tfa}
\bibinfo{author}{\bibfnamefont{G.}~\bibnamefont{Aad}} \bibnamefont{et~al.}
  (\bibinfo{collaboration}{ATLAS Collaboration}), \bibinfo{journal}{Phys.Lett.}
  \textbf{\bibinfo{volume}{B716}}, \bibinfo{pages}{1} (\bibinfo{year}{2012}),
  \eprint{1207.7214}.

\bibitem[{\citenamefont{Chatrchyan et~al.}(2012)}]{Chatrchyan:2012ufa}
\bibinfo{author}{\bibfnamefont{S.}~\bibnamefont{Chatrchyan}}
  \bibnamefont{et~al.} (\bibinfo{collaboration}{CMS Collaboration}),
  \bibinfo{journal}{Phys.Lett.} \textbf{\bibinfo{volume}{B716}},
  \bibinfo{pages}{30} (\bibinfo{year}{2012}), \eprint{1207.7235}.

\bibitem[{CMS-PAS-HIG-13-005()}]{CMS:yva}
CMS-PAS-HIG-13-005 (\bibinfo{year}{2013}).

\bibitem[{\citenamefont{Aad et~al.}(2013)}]{Aad:2013wqa}
\bibinfo{author}{\bibfnamefont{G.}~\bibnamefont{Aad}} \bibnamefont{et~al.}
  (\bibinfo{collaboration}{ATLAS Collaboration}) (\bibinfo{year}{2013}),
  \eprint{1307.1427}.

\bibitem[{\citenamefont{Aaltonen et~al.}(2013)}]{Aaltonen:2013kxa}
\bibinfo{author}{\bibfnamefont{T.}~\bibnamefont{Aaltonen}} \bibnamefont{et~al.}
  (\bibinfo{collaboration}{CDF Collaboration, D0 Collaboration})
  (\bibinfo{year}{2013}), \eprint{1303.6346}.

\bibitem[{\citenamefont{Heinemeyer et~al.}(2013)}]{Heinemeyer:2013tqa}
\bibinfo{author}{\bibfnamefont{S.}~\bibnamefont{Heinemeyer}}
  \bibnamefont{et~al.} (\bibinfo{collaboration}{The LHC Higgs Cross Section
  Working Group}) (\bibinfo{year}{2013}), \eprint{1307.1347}.

\bibitem[{\citenamefont{Frederix et~al.}(2011)\citenamefont{Frederix, Frixione,
  Hirschi, Maltoni, Pittau et~al.}}]{Frederix:2011zi}
\bibinfo{author}{\bibfnamefont{R.}~\bibnamefont{Frederix}},
  \bibinfo{author}{\bibfnamefont{S.}~\bibnamefont{Frixione}},
  \bibinfo{author}{\bibfnamefont{V.}~\bibnamefont{Hirschi}},
  \bibinfo{author}{\bibfnamefont{F.}~\bibnamefont{Maltoni}},
  \bibinfo{author}{\bibfnamefont{R.}~\bibnamefont{Pittau}},
  \bibnamefont{et~al.}, \bibinfo{journal}{Phys.Lett.}
  \textbf{\bibinfo{volume}{B701}}, \bibinfo{pages}{427} (\bibinfo{year}{2011}),
  \eprint{1104.5613}.

\bibitem[{\citenamefont{Degrande et~al.}(2012)\citenamefont{Degrande, Gerard,
  Grojean, Maltoni, and Servant}}]{Degrande:2012gr}
\bibinfo{author}{\bibfnamefont{C.}~\bibnamefont{Degrande}},
  \bibinfo{author}{\bibfnamefont{J.}~\bibnamefont{Gerard}},
  \bibinfo{author}{\bibfnamefont{C.}~\bibnamefont{Grojean}},
  \bibinfo{author}{\bibfnamefont{F.}~\bibnamefont{Maltoni}}, \bibnamefont{and}
  \bibinfo{author}{\bibfnamefont{G.}~\bibnamefont{Servant}},
  \bibinfo{journal}{JHEP} \textbf{\bibinfo{volume}{1207}}, \bibinfo{pages}{036}
  (\bibinfo{year}{2012}), \eprint{1205.1065}.

\bibitem[{\citenamefont{Artoisenet et~al.}(2013)\citenamefont{Artoisenet,
  de~Aquino, Maltoni, and Mattelaer}}]{Artoisenet:2013vfa}
\bibinfo{author}{\bibfnamefont{P.}~\bibnamefont{Artoisenet}},
  \bibinfo{author}{\bibfnamefont{P.}~\bibnamefont{de~Aquino}},
  \bibinfo{author}{\bibfnamefont{F.}~\bibnamefont{Maltoni}}, \bibnamefont{and}
  \bibinfo{author}{\bibfnamefont{O.}~\bibnamefont{Mattelaer}}
  (\bibinfo{year}{2013}), \eprint{1304.6414}.

\bibitem[{\citenamefont{Beenakker et~al.}(2001)\citenamefont{Beenakker,
  Dittmaier, Kramer, Plumper, Spira et~al.}}]{Beenakker:2001rj}
\bibinfo{author}{\bibfnamefont{W.}~\bibnamefont{Beenakker}},
  \bibinfo{author}{\bibfnamefont{S.}~\bibnamefont{Dittmaier}},
  \bibinfo{author}{\bibfnamefont{M.}~\bibnamefont{Kramer}},
  \bibinfo{author}{\bibfnamefont{B.}~\bibnamefont{Plumper}},
  \bibinfo{author}{\bibfnamefont{M.}~\bibnamefont{Spira}},
  \bibnamefont{et~al.}, \bibinfo{journal}{Phys.Rev.Lett.}
  \textbf{\bibinfo{volume}{87}}, \bibinfo{pages}{201805}
  (\bibinfo{year}{2001}), \eprint{hep-ph/0107081}.

\bibitem[{\citenamefont{Beenakker et~al.}(2003)\citenamefont{Beenakker,
  Dittmaier, Kramer, Plumper, Spira et~al.}}]{Beenakker:2002nc}
\bibinfo{author}{\bibfnamefont{W.}~\bibnamefont{Beenakker}},
  \bibinfo{author}{\bibfnamefont{S.}~\bibnamefont{Dittmaier}},
  \bibinfo{author}{\bibfnamefont{M.}~\bibnamefont{Kramer}},
  \bibinfo{author}{\bibfnamefont{B.}~\bibnamefont{Plumper}},
  \bibinfo{author}{\bibfnamefont{M.}~\bibnamefont{Spira}},
  \bibnamefont{et~al.}, \bibinfo{journal}{Nucl.Phys.}
  \textbf{\bibinfo{volume}{B653}}, \bibinfo{pages}{151} (\bibinfo{year}{2003}),
  \eprint{hep-ph/0211352}.

\bibitem[{\citenamefont{Dawson et~al.}(2003{\natexlab{a}})\citenamefont{Dawson,
  Orr, Reina, and Wackeroth}}]{Dawson:2002tg}
\bibinfo{author}{\bibfnamefont{S.}~\bibnamefont{Dawson}},
  \bibinfo{author}{\bibfnamefont{L.}~\bibnamefont{Orr}},
  \bibinfo{author}{\bibfnamefont{L.}~\bibnamefont{Reina}}, \bibnamefont{and}
  \bibinfo{author}{\bibfnamefont{D.}~\bibnamefont{Wackeroth}},
  \bibinfo{journal}{Phys.Rev.} \textbf{\bibinfo{volume}{D67}},
  \bibinfo{pages}{071503} (\bibinfo{year}{2003}{\natexlab{a}}),
  \eprint{hep-ph/0211438}.

\bibitem[{\citenamefont{Dawson et~al.}(2003{\natexlab{b}})\citenamefont{Dawson,
  Jackson, Orr, Reina, and Wackeroth}}]{Dawson:2003zu}
\bibinfo{author}{\bibfnamefont{S.}~\bibnamefont{Dawson}},
  \bibinfo{author}{\bibfnamefont{C.}~\bibnamefont{Jackson}},
  \bibinfo{author}{\bibfnamefont{L.}~\bibnamefont{Orr}},
  \bibinfo{author}{\bibfnamefont{L.}~\bibnamefont{Reina}}, \bibnamefont{and}
  \bibinfo{author}{\bibfnamefont{D.}~\bibnamefont{Wackeroth}},
  \bibinfo{journal}{Phys.Rev.} \textbf{\bibinfo{volume}{D68}},
  \bibinfo{pages}{034022} (\bibinfo{year}{2003}{\natexlab{b}}),
  \eprint{hep-ph/0305087}.

\bibitem[{\citenamefont{Dittmaier et~al.}(2004)\citenamefont{Dittmaier, Kramer,
  and Spira}}]{Dittmaier:2003ej}
\bibinfo{author}{\bibfnamefont{S.}~\bibnamefont{Dittmaier}},
  \bibinfo{author}{\bibfnamefont{.}~\bibnamefont{Kramer},
  \bibfnamefont{Michael}}, \bibnamefont{and}
  \bibinfo{author}{\bibfnamefont{M.}~\bibnamefont{Spira}},
  \bibinfo{journal}{Phys.Rev.} \textbf{\bibinfo{volume}{D70}},
  \bibinfo{pages}{074010} (\bibinfo{year}{2004}), \eprint{hep-ph/0309204}.

\bibitem[{\citenamefont{Plehn et~al.}(2010)\citenamefont{Plehn, Salam, and
  Spannowsky}}]{Plehn:2009rk}
\bibinfo{author}{\bibfnamefont{T.}~\bibnamefont{Plehn}},
  \bibinfo{author}{\bibfnamefont{G.~P.} \bibnamefont{Salam}}, \bibnamefont{and}
  \bibinfo{author}{\bibfnamefont{M.}~\bibnamefont{Spannowsky}},
  \bibinfo{journal}{Phys.Rev.Lett.} \textbf{\bibinfo{volume}{104}},
  \bibinfo{pages}{111801} (\bibinfo{year}{2010}), \eprint{0910.5472}.

\bibitem[{\citenamefont{Mastrolia et~al.}(2012)\citenamefont{Mastrolia,
  Mirabella, and Peraro}}]{Mastrolia:2012bu}
\bibinfo{author}{\bibfnamefont{P.}~\bibnamefont{Mastrolia}},
  \bibinfo{author}{\bibfnamefont{E.}~\bibnamefont{Mirabella}},
  \bibnamefont{and} \bibinfo{author}{\bibfnamefont{T.}~\bibnamefont{Peraro}},
  \bibinfo{journal}{JHEP} \textbf{\bibinfo{volume}{1206}}, \bibinfo{pages}{095}
  (\bibinfo{year}{2012}), \eprint{1203.0291}.

\bibitem[{\citenamefont{Gleisberg et~al.}(2009)\citenamefont{Gleisberg, Hoeche,
  Krauss, Schonherr, Schumann et~al.}}]{Gleisberg:2008ta}
\bibinfo{author}{\bibfnamefont{T.}~\bibnamefont{Gleisberg}},
  \bibinfo{author}{\bibfnamefont{S.}~\bibnamefont{Hoeche}},
  \bibinfo{author}{\bibfnamefont{F.}~\bibnamefont{Krauss}},
  \bibinfo{author}{\bibfnamefont{M.}~\bibnamefont{Schonherr}},
  \bibinfo{author}{\bibfnamefont{S.}~\bibnamefont{Schumann}},
  \bibnamefont{et~al.}, \bibinfo{journal}{JHEP}
  \textbf{\bibinfo{volume}{0902}}, \bibinfo{pages}{007} (\bibinfo{year}{2009}),
  \eprint{0811.4622}.

\bibitem[{\citenamefont{Krauss et~al.}(2002)\citenamefont{Krauss, Kuhn, and
  Soff}}]{Krauss:2001iv}
\bibinfo{author}{\bibfnamefont{F.}~\bibnamefont{Krauss}},
  \bibinfo{author}{\bibfnamefont{R.}~\bibnamefont{Kuhn}}, \bibnamefont{and}
  \bibinfo{author}{\bibfnamefont{G.}~\bibnamefont{Soff}},
  \bibinfo{journal}{JHEP} \textbf{\bibinfo{volume}{0202}}, \bibinfo{pages}{044}
  (\bibinfo{year}{2002}), \eprint{hep-ph/0109036}.

\bibitem[{\citenamefont{Catani et~al.}(2002)\citenamefont{Catani, Dittmaier,
  Seymour, and Trocsanyi}}]{Catani:2002hc}
\bibinfo{author}{\bibfnamefont{S.}~\bibnamefont{Catani}},
  \bibinfo{author}{\bibfnamefont{S.}~\bibnamefont{Dittmaier}},
  \bibinfo{author}{\bibfnamefont{M.~H.} \bibnamefont{Seymour}},
  \bibnamefont{and}
  \bibinfo{author}{\bibfnamefont{Z.}~\bibnamefont{Trocsanyi}},
  \bibinfo{journal}{Nucl.Phys.} \textbf{\bibinfo{volume}{B627}},
  \bibinfo{pages}{189} (\bibinfo{year}{2002}), \eprint{hep-ph/0201036}.

\bibitem[{\citenamefont{Gleisberg and Krauss}(2008)}]{Gleisberg:2007md}
\bibinfo{author}{\bibfnamefont{T.}~\bibnamefont{Gleisberg}} \bibnamefont{and}
  \bibinfo{author}{\bibfnamefont{F.}~\bibnamefont{Krauss}},
  \bibinfo{journal}{Eur.Phys.J.} \textbf{\bibinfo{volume}{C53}},
  \bibinfo{pages}{501} (\bibinfo{year}{2008}), \eprint{0709.2881}.

\bibitem[{\citenamefont{Cullen et~al.}(2012)\citenamefont{Cullen, Greiner,
  Heinrich, Luisoni, Mastrolia et~al.}}]{Cullen:2011ac}
\bibinfo{author}{\bibfnamefont{G.}~\bibnamefont{Cullen}},
  \bibinfo{author}{\bibfnamefont{N.}~\bibnamefont{Greiner}},
  \bibinfo{author}{\bibfnamefont{G.}~\bibnamefont{Heinrich}},
  \bibinfo{author}{\bibfnamefont{G.}~\bibnamefont{Luisoni}},
  \bibinfo{author}{\bibfnamefont{P.}~\bibnamefont{Mastrolia}},
  \bibnamefont{et~al.}, \bibinfo{journal}{Eur.Phys.J.}
  \textbf{\bibinfo{volume}{C72}}, \bibinfo{pages}{1889} (\bibinfo{year}{2012}),
  \eprint{1111.2034}.

\bibitem[{\citenamefont{Nogueira}(1993)}]{Nogueira:1991ex}
\bibinfo{author}{\bibfnamefont{P.}~\bibnamefont{Nogueira}},
  \bibinfo{journal}{J.Comput.Phys.} \textbf{\bibinfo{volume}{105}},
  \bibinfo{pages}{279} (\bibinfo{year}{1993}).

\bibitem[{\citenamefont{Vermaseren}(2000)}]{Vermaseren:2000nd}
\bibinfo{author}{\bibfnamefont{J.~A.~M.} \bibnamefont{Vermaseren}}
  (\bibinfo{year}{2000}), \eprint{math-ph/0010025}.

\bibitem[{\citenamefont{Reiter}(2010)}]{Reiter:2009ts}
\bibinfo{author}{\bibfnamefont{T.}~\bibnamefont{Reiter}},
  \bibinfo{journal}{Comput.Phys.Commun.} \textbf{\bibinfo{volume}{181}},
  \bibinfo{pages}{1301} (\bibinfo{year}{2010}), \eprint{0907.3714}.

\bibitem[{\citenamefont{Cullen et~al.}(2011{\natexlab{a}})\citenamefont{Cullen,
  Koch-Janusz, and Reiter}}]{Cullen:2010jv}
\bibinfo{author}{\bibfnamefont{G.}~\bibnamefont{Cullen}},
  \bibinfo{author}{\bibfnamefont{M.}~\bibnamefont{Koch-Janusz}},
  \bibnamefont{and} \bibinfo{author}{\bibfnamefont{T.}~\bibnamefont{Reiter}},
  \bibinfo{journal}{Comput.Phys.Commun.} \textbf{\bibinfo{volume}{182}},
  \bibinfo{pages}{2368} (\bibinfo{year}{2011}{\natexlab{a}}),
  \eprint{1008.0803}.

\bibitem[{\citenamefont{Kuipers et~al.}(2013)\citenamefont{Kuipers, Ueda,
  Vermaseren, and Vollinga}}]{Kuipers:2012rf}
\bibinfo{author}{\bibfnamefont{J.}~\bibnamefont{Kuipers}},
  \bibinfo{author}{\bibfnamefont{T.}~\bibnamefont{Ueda}},
  \bibinfo{author}{\bibfnamefont{J.}~\bibnamefont{Vermaseren}},
  \bibnamefont{and} \bibinfo{author}{\bibfnamefont{J.}~\bibnamefont{Vollinga}},
  \bibinfo{journal}{Comput.Phys.Commun.} \textbf{\bibinfo{volume}{184}},
  \bibinfo{pages}{1453} (\bibinfo{year}{2013}), \eprint{1203.6543}.

\bibitem[{\citenamefont{Ossola et~al.}(2007{\natexlab{a}})\citenamefont{Ossola,
  Papadopoulos, and Pittau}}]{Ossola:2006us}
\bibinfo{author}{\bibfnamefont{G.}~\bibnamefont{Ossola}},
  \bibinfo{author}{\bibfnamefont{C.~G.} \bibnamefont{Papadopoulos}},
  \bibnamefont{and} \bibinfo{author}{\bibfnamefont{R.}~\bibnamefont{Pittau}},
  \bibinfo{journal}{Nucl.Phys.} \textbf{\bibinfo{volume}{B763}},
  \bibinfo{pages}{147} (\bibinfo{year}{2007}{\natexlab{a}}),
  \eprint{hep-ph/0609007}.

\bibitem[{\citenamefont{Ossola et~al.}(2007{\natexlab{b}})\citenamefont{Ossola,
  Papadopoulos, and Pittau}}]{Ossola:2007bb}
\bibinfo{author}{\bibfnamefont{G.}~\bibnamefont{Ossola}},
  \bibinfo{author}{\bibfnamefont{C.~G.} \bibnamefont{Papadopoulos}},
  \bibnamefont{and} \bibinfo{author}{\bibfnamefont{R.}~\bibnamefont{Pittau}},
  \bibinfo{journal}{JHEP} \textbf{\bibinfo{volume}{0707}}, \bibinfo{pages}{085}
  (\bibinfo{year}{2007}{\natexlab{b}}), \eprint{0704.1271}.

\bibitem[{\citenamefont{Ellis et~al.}(2008)\citenamefont{Ellis, Giele, and
  Kunszt}}]{Ellis:2007br}
\bibinfo{author}{\bibfnamefont{R.~K.} \bibnamefont{Ellis}},
  \bibinfo{author}{\bibfnamefont{W.~T.} \bibnamefont{Giele}}, \bibnamefont{and}
  \bibinfo{author}{\bibfnamefont{Z.}~\bibnamefont{Kunszt}},
  \bibinfo{journal}{JHEP} \textbf{\bibinfo{volume}{03}}, \bibinfo{pages}{003}
  (\bibinfo{year}{2008}), \eprint{0708.2398}.

\bibitem[{\citenamefont{Ossola et~al.}(2008)\citenamefont{Ossola, Papadopoulos,
  and Pittau}}]{Ossola:2008xq}
\bibinfo{author}{\bibfnamefont{G.}~\bibnamefont{Ossola}},
  \bibinfo{author}{\bibfnamefont{C.~G.} \bibnamefont{Papadopoulos}},
  \bibnamefont{and} \bibinfo{author}{\bibfnamefont{R.}~\bibnamefont{Pittau}},
  \bibinfo{journal}{JHEP} \textbf{\bibinfo{volume}{0805}}, \bibinfo{pages}{004}
  (\bibinfo{year}{2008}), \eprint{0802.1876}.

\bibitem[{\citenamefont{Mastrolia et~al.}(2008)\citenamefont{Mastrolia, Ossola,
  Papadopoulos, and Pittau}}]{Mastrolia:2008jb}
\bibinfo{author}{\bibfnamefont{P.}~\bibnamefont{Mastrolia}},
  \bibinfo{author}{\bibfnamefont{G.}~\bibnamefont{Ossola}},
  \bibinfo{author}{\bibfnamefont{C.}~\bibnamefont{Papadopoulos}},
  \bibnamefont{and} \bibinfo{author}{\bibfnamefont{R.}~\bibnamefont{Pittau}},
  \bibinfo{journal}{JHEP} \textbf{\bibinfo{volume}{0806}}, \bibinfo{pages}{030}
  (\bibinfo{year}{2008}), \eprint{0803.3964}.

\bibitem[{\citenamefont{Mastrolia et~al.}(2010)\citenamefont{Mastrolia, Ossola,
  Reiter, and Tramontano}}]{Mastrolia:2010nb}
\bibinfo{author}{\bibfnamefont{P.}~\bibnamefont{Mastrolia}},
  \bibinfo{author}{\bibfnamefont{G.}~\bibnamefont{Ossola}},
  \bibinfo{author}{\bibfnamefont{T.}~\bibnamefont{Reiter}}, \bibnamefont{and}
  \bibinfo{author}{\bibfnamefont{F.}~\bibnamefont{Tramontano}},
  \bibinfo{journal}{JHEP} \textbf{\bibinfo{volume}{1008}}, \bibinfo{pages}{080}
  (\bibinfo{year}{2010}), \eprint{1006.0710}.

\bibitem[{\citenamefont{Heinrich et~al.}(2010)\citenamefont{Heinrich, Ossola,
  Reiter, and Tramontano}}]{Heinrich:2010ax}
\bibinfo{author}{\bibfnamefont{G.}~\bibnamefont{Heinrich}},
  \bibinfo{author}{\bibfnamefont{G.}~\bibnamefont{Ossola}},
  \bibinfo{author}{\bibfnamefont{T.}~\bibnamefont{Reiter}}, \bibnamefont{and}
  \bibinfo{author}{\bibfnamefont{F.}~\bibnamefont{Tramontano}},
  \bibinfo{journal}{JHEP} \textbf{\bibinfo{volume}{1010}}, \bibinfo{pages}{105}
  (\bibinfo{year}{2010}), \eprint{1008.2441}.

\bibitem[{\citenamefont{van Hameren}(2011)}]{vanHameren:2010cp}
\bibinfo{author}{\bibfnamefont{A.}~\bibnamefont{van Hameren}},
  \bibinfo{journal}{Comput.Phys.Commun.} \textbf{\bibinfo{volume}{182}},
  \bibinfo{pages}{2427} (\bibinfo{year}{2011}), \eprint{1007.4716}.

\bibitem[{\citenamefont{Binoth et~al.}(2010)\citenamefont{Binoth, Boudjema,
  Dissertori, Lazopoulos, Denner et~al.}}]{Binoth:2010xt}
\bibinfo{author}{\bibfnamefont{T.}~\bibnamefont{Binoth}},
  \bibinfo{author}{\bibfnamefont{F.}~\bibnamefont{Boudjema}},
  \bibinfo{author}{\bibfnamefont{G.}~\bibnamefont{Dissertori}},
  \bibinfo{author}{\bibfnamefont{A.}~\bibnamefont{Lazopoulos}},
  \bibinfo{author}{\bibfnamefont{A.}~\bibnamefont{Denner}},
  \bibnamefont{et~al.}, \bibinfo{journal}{Comput.Phys.Commun.}
  \textbf{\bibinfo{volume}{181}}, \bibinfo{pages}{1612} (\bibinfo{year}{2010}),
  \eprint{1001.1307}.

\bibitem[{\citenamefont{van Deurzen et~al.}(2013)\citenamefont{van Deurzen,
  Greiner, Luisoni, Mastrolia, Mirabella et~al.}}]{vanDeurzen:2013rv}
\bibinfo{author}{\bibfnamefont{H.}~\bibnamefont{van Deurzen}},
  \bibinfo{author}{\bibfnamefont{N.}~\bibnamefont{Greiner}},
  \bibinfo{author}{\bibfnamefont{G.}~\bibnamefont{Luisoni}},
  \bibinfo{author}{\bibfnamefont{P.}~\bibnamefont{Mastrolia}},
  \bibinfo{author}{\bibfnamefont{E.}~\bibnamefont{Mirabella}},
  \bibnamefont{et~al.}, \bibinfo{journal}{Phys.Lett.}
  \textbf{\bibinfo{volume}{B721}}, \bibinfo{pages}{74} (\bibinfo{year}{2013}),
  \eprint{1301.0493}.

\bibitem[{\citenamefont{Cullen et~al.}(2013)\citenamefont{Cullen, van Deurzen,
  Greiner, Luisoni, Mastrolia et~al.}}]{Cullen:2013saa}
\bibinfo{author}{\bibfnamefont{G.}~\bibnamefont{Cullen}},
  \bibinfo{author}{\bibfnamefont{H.}~\bibnamefont{van Deurzen}},
  \bibinfo{author}{\bibfnamefont{N.}~\bibnamefont{Greiner}},
  \bibinfo{author}{\bibfnamefont{G.}~\bibnamefont{Luisoni}},
  \bibinfo{author}{\bibfnamefont{P.}~\bibnamefont{Mastrolia}},
  \bibnamefont{et~al.}, \bibinfo{journal}{Phys.Rev.Lett.}
  \textbf{\bibinfo{volume}{111}}, \bibinfo{pages}{131801}
  (\bibinfo{year}{2013}), \eprint{1307.4737}.

\bibitem[{\citenamefont{Maltoni and Stelzer}(2003)}]{Maltoni:2002qb}
\bibinfo{author}{\bibfnamefont{F.}~\bibnamefont{Maltoni}} \bibnamefont{and}
  \bibinfo{author}{\bibfnamefont{T.}~\bibnamefont{Stelzer}},
  \bibinfo{journal}{JHEP} \textbf{\bibinfo{volume}{0302}}, \bibinfo{pages}{027}
  (\bibinfo{year}{2003}), \eprint{hep-ph/0208156}.

\bibitem[{\citenamefont{Frederix et~al.}(2008)\citenamefont{Frederix, Gehrmann,
  and Greiner}}]{Frederix:2008hu}
\bibinfo{author}{\bibfnamefont{R.}~\bibnamefont{Frederix}},
  \bibinfo{author}{\bibfnamefont{T.}~\bibnamefont{Gehrmann}}, \bibnamefont{and}
  \bibinfo{author}{\bibfnamefont{N.}~\bibnamefont{Greiner}},
  \bibinfo{journal}{JHEP} \textbf{\bibinfo{volume}{0809}}, \bibinfo{pages}{122}
  (\bibinfo{year}{2008}), \eprint{0808.2128}.

\bibitem[{\citenamefont{Frederix et~al.}(2010)\citenamefont{Frederix, Gehrmann,
  and Greiner}}]{Frederix:2010cj}
\bibinfo{author}{\bibfnamefont{R.}~\bibnamefont{Frederix}},
  \bibinfo{author}{\bibfnamefont{T.}~\bibnamefont{Gehrmann}}, \bibnamefont{and}
  \bibinfo{author}{\bibfnamefont{N.}~\bibnamefont{Greiner}},
  \bibinfo{journal}{JHEP} \textbf{\bibinfo{volume}{1006}}, \bibinfo{pages}{086}
  (\bibinfo{year}{2010}), \eprint{1004.2905}.

\bibitem[{\citenamefont{Hoeche et~al.}(2013)\citenamefont{Hoeche, Huang,
  Luisoni, Schoenherr, and Winter}}]{Hoeche:2013mua}
\bibinfo{author}{\bibfnamefont{S.}~\bibnamefont{Hoeche}},
  \bibinfo{author}{\bibfnamefont{J.}~\bibnamefont{Huang}},
  \bibinfo{author}{\bibfnamefont{G.}~\bibnamefont{Luisoni}},
  \bibinfo{author}{\bibfnamefont{M.}~\bibnamefont{Schoenherr}},
  \bibnamefont{and} \bibinfo{author}{\bibfnamefont{J.}~\bibnamefont{Winter}},
  \bibinfo{journal}{Phys.Rev.} \textbf{\bibinfo{volume}{D88}},
  \bibinfo{pages}{014040} (\bibinfo{year}{2013}), \eprint{1306.2703}.

\bibitem[{\citenamefont{Forde}(2007)}]{Forde:2007mi}
\bibinfo{author}{\bibfnamefont{D.}~\bibnamefont{Forde}},
  \bibinfo{journal}{Phys. Rev.} \textbf{\bibinfo{volume}{D75}},
  \bibinfo{pages}{125019} (\bibinfo{year}{2007}), \eprint{0704.1835}.

\bibitem[{\citenamefont{Kilgore}(2007)}]{Kilgore:2007qr}
\bibinfo{author}{\bibfnamefont{W.~B.} \bibnamefont{Kilgore}}
  (\bibinfo{year}{2007}), \eprint{0711.5015}.

\bibitem[{\citenamefont{Badger}(2009)}]{Badger:2008cm}
\bibinfo{author}{\bibfnamefont{S.~D.} \bibnamefont{Badger}},
  \bibinfo{journal}{JHEP} \textbf{\bibinfo{volume}{01}}, \bibinfo{pages}{049}
  (\bibinfo{year}{2009}), \eprint{0806.4600}.

\bibitem[{\citenamefont{Catani et~al.}(2001)\citenamefont{Catani, Dittmaier,
  and Trocsanyi}}]{Catani:2000ef}
\bibinfo{author}{\bibfnamefont{S.}~\bibnamefont{Catani}},
  \bibinfo{author}{\bibfnamefont{S.}~\bibnamefont{Dittmaier}},
  \bibnamefont{and}
  \bibinfo{author}{\bibfnamefont{Z.}~\bibnamefont{Trocsanyi}},
  \bibinfo{journal}{Phys.Lett.} \textbf{\bibinfo{volume}{B500}},
  \bibinfo{pages}{149} (\bibinfo{year}{2001}), \eprint{hep-ph/0011222}.

\bibitem[{\citenamefont{Binoth et~al.}(2009)\citenamefont{Binoth, Guillet,
  Heinrich, Pilon, and Reiter}}]{Binoth:2008uq}
\bibinfo{author}{\bibfnamefont{T.}~\bibnamefont{Binoth}},
  \bibinfo{author}{\bibfnamefont{J.-P.} \bibnamefont{Guillet}},
  \bibinfo{author}{\bibfnamefont{G.}~\bibnamefont{Heinrich}},
  \bibinfo{author}{\bibfnamefont{E.}~\bibnamefont{Pilon}}, \bibnamefont{and}
  \bibinfo{author}{\bibfnamefont{T.}~\bibnamefont{Reiter}},
  \bibinfo{journal}{Comput.Phys.Commun.} \textbf{\bibinfo{volume}{180}},
  \bibinfo{pages}{2317} (\bibinfo{year}{2009}), \eprint{0810.0992}.

\bibitem[{\citenamefont{Cullen et~al.}(2011{\natexlab{b}})\citenamefont{Cullen,
  Guillet, Heinrich, Kleinschmidt, Pilon et~al.}}]{Cullen:2011kv}
\bibinfo{author}{\bibfnamefont{G.}~\bibnamefont{Cullen}},
  \bibinfo{author}{\bibfnamefont{J.}~\bibnamefont{Guillet}},
  \bibinfo{author}{\bibfnamefont{G.}~\bibnamefont{Heinrich}},
  \bibinfo{author}{\bibfnamefont{T.}~\bibnamefont{Kleinschmidt}},
  \bibinfo{author}{\bibfnamefont{E.}~\bibnamefont{Pilon}},
  \bibnamefont{et~al.}, \bibinfo{journal}{Comput.Phys.Commun.}
  \textbf{\bibinfo{volume}{182}}, \bibinfo{pages}{2276}
  (\bibinfo{year}{2011}{\natexlab{b}}), \eprint{1101.5595}.

\bibitem[{\citenamefont{Hirschi et~al.}(2011)\citenamefont{Hirschi, Frederix,
  Frixione, Garzelli, Maltoni et~al.}}]{Hirschi:2011pa}
\bibinfo{author}{\bibfnamefont{V.}~\bibnamefont{Hirschi}},
  \bibinfo{author}{\bibfnamefont{R.}~\bibnamefont{Frederix}},
  \bibinfo{author}{\bibfnamefont{S.}~\bibnamefont{Frixione}},
  \bibinfo{author}{\bibfnamefont{M.~V.} \bibnamefont{Garzelli}},
  \bibinfo{author}{\bibfnamefont{F.}~\bibnamefont{Maltoni}},
  \bibnamefont{et~al.}, \bibinfo{journal}{JHEP}
  \textbf{\bibinfo{volume}{1105}}, \bibinfo{pages}{044} (\bibinfo{year}{2011}),
  \eprint{1103.0621}.

\bibitem[{\citenamefont{Cacciari and Salam}(2006)}]{Cacciari:2005hq}
\bibinfo{author}{\bibfnamefont{M.}~\bibnamefont{Cacciari}} \bibnamefont{and}
  \bibinfo{author}{\bibfnamefont{G.~P.} \bibnamefont{Salam}},
  \bibinfo{journal}{Phys.Lett.} \textbf{\bibinfo{volume}{B641}},
  \bibinfo{pages}{57} (\bibinfo{year}{2006}), \eprint{hep-ph/0512210}.

\bibitem[{\citenamefont{Cacciari et~al.}(2008)\citenamefont{Cacciari, Salam,
  and Soyez}}]{Cacciari:2008gp}
\bibinfo{author}{\bibfnamefont{M.}~\bibnamefont{Cacciari}},
  \bibinfo{author}{\bibfnamefont{G.~P.} \bibnamefont{Salam}}, \bibnamefont{and}
  \bibinfo{author}{\bibfnamefont{G.}~\bibnamefont{Soyez}},
  \bibinfo{journal}{JHEP} \textbf{\bibinfo{volume}{0804}}, \bibinfo{pages}{063}
  (\bibinfo{year}{2008}), \eprint{0802.1189}.

\bibitem[{\citenamefont{Cacciari et~al.}(2012)\citenamefont{Cacciari, Salam,
  and Soyez}}]{Cacciari:2011ma}
\bibinfo{author}{\bibfnamefont{M.}~\bibnamefont{Cacciari}},
  \bibinfo{author}{\bibfnamefont{G.~P.} \bibnamefont{Salam}}, \bibnamefont{and}
  \bibinfo{author}{\bibfnamefont{G.}~\bibnamefont{Soyez}},
  \bibinfo{journal}{Eur.Phys.J.} \textbf{\bibinfo{volume}{C72}},
  \bibinfo{pages}{1896} (\bibinfo{year}{2012}), \eprint{1111.6097}.

\bibitem[{\citenamefont{Pumplin et~al.}(2002)\citenamefont{Pumplin, Stump,
  Huston, Lai, Nadolsky et~al.}}]{Pumplin:2002vw}
\bibinfo{author}{\bibfnamefont{J.}~\bibnamefont{Pumplin}},
  \bibinfo{author}{\bibfnamefont{D.}~\bibnamefont{Stump}},
  \bibinfo{author}{\bibfnamefont{J.}~\bibnamefont{Huston}},
  \bibinfo{author}{\bibfnamefont{H.}~\bibnamefont{Lai}},
  \bibinfo{author}{\bibfnamefont{P.~M.} \bibnamefont{Nadolsky}},
  \bibnamefont{et~al.}, \bibinfo{journal}{JHEP}
  \textbf{\bibinfo{volume}{0207}}, \bibinfo{pages}{012} (\bibinfo{year}{2002}),
  \eprint{hep-ph/0201195}.

\bibitem[{\citenamefont{Lai et~al.}(2010)\citenamefont{Lai, Guzzi, Huston, Li,
  Nadolsky et~al.}}]{Lai:2010vv}
\bibinfo{author}{\bibfnamefont{H.-L.} \bibnamefont{Lai}},
  \bibinfo{author}{\bibfnamefont{M.}~\bibnamefont{Guzzi}},
  \bibinfo{author}{\bibfnamefont{J.}~\bibnamefont{Huston}},
  \bibinfo{author}{\bibfnamefont{Z.}~\bibnamefont{Li}},
  \bibinfo{author}{\bibfnamefont{P.~M.} \bibnamefont{Nadolsky}},
  \bibnamefont{et~al.}, \bibinfo{journal}{Phys.Rev.}
  \textbf{\bibinfo{volume}{D82}}, \bibinfo{pages}{074024}
  (\bibinfo{year}{2010}), \eprint{1007.2241}.

\bibitem[{\citenamefont{Butterworth et~al.}(2008)\citenamefont{Butterworth,
  Davison, Rubin, and Salam}}]{Butterworth:2008iy}
\bibinfo{author}{\bibfnamefont{J.~M.} \bibnamefont{Butterworth}},
  \bibinfo{author}{\bibfnamefont{A.~R.} \bibnamefont{Davison}},
  \bibinfo{author}{\bibfnamefont{M.}~\bibnamefont{Rubin}}, \bibnamefont{and}
  \bibinfo{author}{\bibfnamefont{G.~P.} \bibnamefont{Salam}},
  \bibinfo{journal}{Phys.Rev.Lett.} \textbf{\bibinfo{volume}{100}},
  \bibinfo{pages}{242001} (\bibinfo{year}{2008}), \eprint{0802.2470}.

\end{thebibliography}

\end{document}